\documentclass[reprint,prl,superscriptaddress,nofootinbib]{revtex4-1}

\usepackage{graphicx}
\usepackage{dcolumn}
\usepackage{bm}
\usepackage{float}
\usepackage{amsthm,amssymb,amsmath,url,braket} 
\usepackage{threeparttable}
\usepackage{cancel}
\usepackage{color}

\graphicspath {{figure/}}

\begin{document}

\title{Symmetry of Berry and spin Berry curvatures in ferromagnetic CoPt}

\author{Guanxiong Qu}
\affiliation{Department of Physics, The University of Tokyo, 7-3-1 Hongo, Bunkyo 113-0033, Japan}

\author{Kohji Nakamura}
\affiliation{Department of Physics Engineering, Mie University, Tsu, Mie 514-8507, Japan}

\author{Masamitsu Hayashi}
\affiliation{Department of Physics, The University of Tokyo, 7-3-1 Hongo, Bunkyo 113-0033, Japan}
\affiliation{National Institute for Materials Science, 1-2-1 Sengen, Tsukuba 305-0047, Japan}


\date{\today}

\begin{abstract}
The intrinsic spin Hall conductivity and the anomalous Hall conductivity of ferromagnetic L1$_0$-CoPt are studied using first principle calculations of the spin Berry and Berry curvatures, respectively. We find that the Berry curvature and the spin Berry curvature exhibit different symmetry with respect to that of the band structure. 
The Berry curvature preserves the $C_{4v}$ crystal rotation symmetry along the c-axis whereas the symmetry of the spin Berry curvature reduces to $C_{2v}$. Contributions to the Berry curvature and the spin Berry curvature are classified by the spin character of bands crossing the Fermi level. We find that the reduced symmetry of the spin Berry curvature is due to band crossing points with opposite spin characters. From model Hamiltonian analyses, we show the universality of this distinct symmetry reduction of the spin Berry curvature with respect to the Berry curvature: it can be accounted for based on the form of spin current operator and velocity operator in the Kubo formula. Finally, we discuss the consequence of the reduced symmetry of the spin Berry curvature on the relationship between the anomalous Hall and spin Hall conductivity. When band crossing points with opposite spin characters are present in the reciprocal space, which is often the case, the anomalous Hall conductivity does not simply scale with the spin Hall conductivity with the scaling factor being the spin polarization at the Fermi level. 
\end{abstract}


\maketitle


The spin Hall effect (SHE) allows generation of spin current transverse to the current flow direction\cite{Dyakonov1971,Hirsch1999,Murakami2004,Sinova2004}. It was first experimentally found in semiconductors \cite{,Kato2004} and later in paramagnetic heavy metals (e.g. Pt, W)\cite{Liu2012,Pai2012}, where the effect is typically larger in the latter than the former. The large spin current can diffuse into adjacent ferromagnetic layer(s) and cause excitation of magnetic moments via the spin transfer torque. A large effort has been put forward to explore materials with large SHE. One of the material systems that has gained interest recently is the ferromagnetic metals wherein considerable amount of current induced spin current has been reported\cite{Tian2016,Das2017,Iihama2018,Beak2018} despite its relatively small spin orbit coupling. It is thus of high importance to understand the mechanism of spin current generation in ferromagnets.

Ferromagnets are distinguished by their large exchange interaction, or macroscopically by the ferromagnetic ordering. Electron transport in ferromagnets is significantly influenced by the ordering: the anisotropic magnetoresistance\cite{McGuire1975} and the anomalous Hall effect (AHE)\cite{Nagaosa2010} are two well known examples. Theoretical studies on the AHE have evolved, from classical treatment of the electron transport to microscopic quantum models, including extrinsic and intrinsic contributions. In modern formalism the intrinsic AHE is directly related to non-vanishing Berry curvature in the momentum space. Practically, the intrinsic contribution of the anomalous Hall conductivity (AHC) can be calculated from the Kubo formula in the spectral representation\cite{Yao2004}, which is mathematically equivalent to calculating the Berry curvature. In analogy to AHC, the intrinsic spin Hall conductivity (SHC) can also be calculated through the Kubo formula by properly defining the spin current operator\cite{Sun2005,Shi2006,Murakami2006,Guo2008}. It is now commonly understood that the AHE and the SHE can be described using the same theoretical framework\cite{Sinova2015}. Experiments show that the intrinsic contribution to the AHC and SHC are dominant in moderately dirty systems\cite{miyasato2007ahe,onoda2008ahe,sagasta2016tuning}. Nevertheless, limited number of studies have been reported to discuss the AHE and SHE together to provide direct comparison\cite{Zhang2017,Zelezny2017}.
In this context, ferromagnets are undoubtably the best playground to study both effects.

Here we use L1$_0$-CoPt as a prototype of ferromagnets to study the intrinsic AHC and SHC via calculations of the Berry and spin Berry curvatures. Particular emphasis is put on the symmetry of Berry and spin Berry curvatures with respect to the crystal symmetry. First-principle calculations based on density functional theory (DFT) and the Kubo formula are used to calculate the Berry and spin Berry curvatures projected in the momentum space to show their global symmetries. We find the Berry curvature preserves the crystal symmetry whereas the spin Berry curvature has a reduced symmetry. 
We construct a simple model Hamiltonian to demonstrate the universality of the symmetry reduction of the spin Berry curvatures. 

\begin{figure}[b]
\includegraphics[width=8cm]{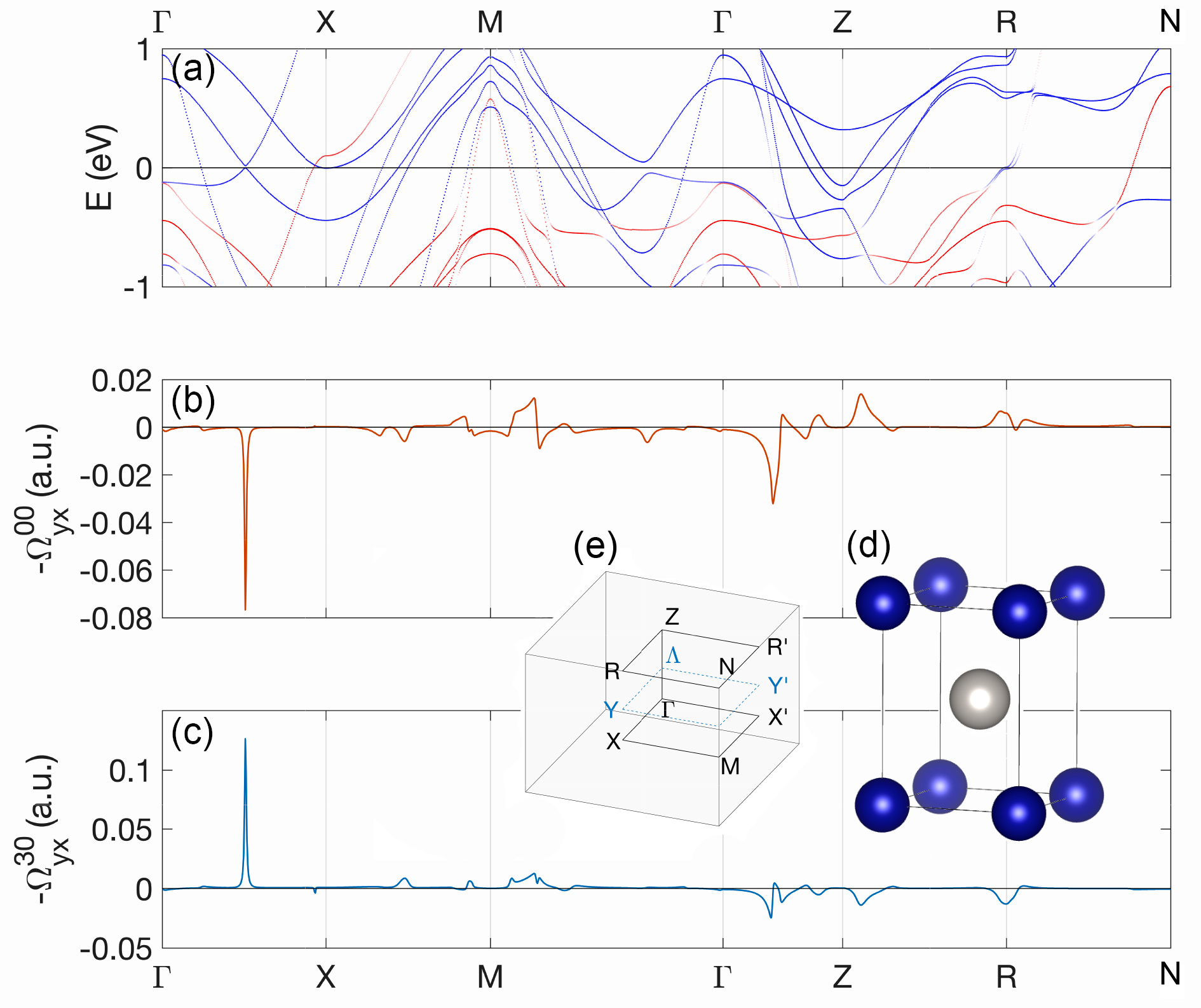}%
\caption{\label{fig:2} (a) Band structure, (b) Berry curvature, and (c) spin Berry curvature of L1$_0$-CoPt along high symmetry k-path. The bands are colored by its spin characters with majority (red) and minority (blue).  (d) Primitive cell of L1$_0$-CoPt. (e) The first Brilloiun zone and selected high symmetry $\bm{k}$ points.} 
\label{fig:1}
\end{figure}

DFT calculations are performed using the full-potential linearized augmented-plane-wave method (FLAPW)\cite{wimmer1981wimmer,weinert1982total,nakamura2003enhancement} with generalized gradient approximation (GGA)\cite{perdew1996generalized} for exchange correlation. The primitive cell of  L1$_0$-CoPt, Fig.\ref{fig:1}(d), is constructed with experimental lattice constant
$c=3.71 \text{\AA}, a=2.69 \text{\AA}$.
Spin orbit coupling (SOC) is treated through a second variational method\cite{PhysRevB.42.5433} and zero temperature is assumed.  

The intrinsic AHC and SHC are obtained from the linear response Kubo formula in the static limit \cite{fang2003anomalous,yao2004first,PhysRevLett.92.126603,PhysRevLett.100.096401,FENG2016428}. We define a general velocity operator $v_i^{\alpha}= \frac{1}{2} \{ \sigma^{\alpha}, v_i\}$ with subscript $i(=x,y,z)$ indicating the spatial coordinate. The greek index $\alpha$ has four components ($\alpha=0,1,2,3$)\cite{jin20062}. $\sigma^{\alpha}$ is defined as $\sigma^{\alpha}=( I_2,\sigma_1,\sigma_2,\sigma_3)$, where $I$ is a $2 \times 2$ unit matrix and $\sigma_l$ is the Pauli matrix with $l=1,2,3$. $\hat{v}_i^{0}$ and $\hat{v}_i^{l}$ are the velocity operator and the spin current operator, respectively. Note that $i$ and $l$ in $v_i^{l}$ represent the flow and spin directions of the spin current, respectively.
Under these conventions, the Berry curvature ($\Omega_{ij}^{00}$) and the spin Berry curvature ($\Omega_{ij}^{l0}$) are expressed as,
\begin{equation}
\begin{aligned}
\Omega_{ij}^{\alpha 0}( \bm{k} ) =& -\sum_{n' \neq n} \Big[ f(\epsilon_n(\bm{k})) -f(\epsilon_{n'}(\bm{k})) \Big]  \\
& \times \frac{ \text{Im} \Big[ \bra{\bm{k},n} \hat{v}_i^{\alpha} \ket{k,n'} \bra{ \bm{k} ,n'} \hat{v}_j^{0} \ket{ \bm{k} ,n} \Big] }{\Big(\epsilon_{n}(\bm{k}) - \epsilon_{n'}(\bm{k}) \Big)^2} 
\label{eq:1}
\end{aligned}
\end{equation}
where $\ket{\bm{k},n}$ is the Bloch state with energy $\epsilon_{n}(\bm{k})$ and wave vector $\bm{k}$; $n$ describes the band index. $\hat{v}_i^{\alpha}$ is the generalized velocity operator ($i=x,y,z$ and $\alpha = 0,1,2,3$) and $f(\epsilon)$ is the Fermi distribution function. The off-diagonal conductivity tensor ($\sigma_{ij}^{\alpha 0}$) can be obtained by integrating the Berry and spin Berry curvatures in the first Brillouin zone (BZ):
\begin{equation}
\sigma_{ij}^{\alpha 0} = - \dfrac{e^2}{\hbar} \int_{BZ} \frac{d^3 \bm{k} }{(2 \pi)^3} \Omega_{ij}^{\alpha 0}( \bm{k} )
\label{eq:2}
\end{equation}
Here, $e$ and $\hbar$ are the electric charge and the reduced Planck constant, respectively, $\sigma_{ij}^{00}$ represents the AHC and $\sigma_{ij}^{l0}$ represents the SHC with the spin quantization axis along the $l$-axis in the Cartesian space. Unless noted otherwise, we discuss the SHC with the spin quantization axis always along the $z$ axis ($\sigma_{ij}^{30}$). 
The calculated $\sigma_{yx}^{00}$ and $\sigma_{yx}^{30}$ from the first principle calculations are $-3$ S/cm and $787$ S/cm, respectively.

Figures ~\ref{fig:1}(a), (b) and (c) show, respectively, the band structure, the non-vanishing component of the Berry curvature ($\Omega_{yx}^{00}$) and the spin Berry curvature ($\Omega_{yx}^{30}$) along the high symmetry wave vector ($\bm{k}$)-path surrounding the irreducible first BZ. 
Large contributions to the $\Omega_{yx}^{00}$ and $\Omega_{yx}^{30}$ are found at the boundary of the $\Gamma$-X-M plane: a major peak of $\Omega_{yx}^{00}$ and $\Omega_{yx}^{30}$ is found along the $\Gamma$-X line. The $\bm{k}$ vectors where such peaks (or valleys) occur tend to coincide with the position where two bands cross the Fermi level. To analyze the correlation of Berry and spin Berry curvatures, we study $\Omega_{yx}^{00}$ and $\Omega_{yx}^{30}$ mapped in the (001)-plane of the momentum space, as shown in Figs.~\ref{fig:2}(a,b). For sake of discussion, we categorize the band crossings points into two classes: the corresponding bands have I. the same spin character; II. the opposite spin character. 

\begin{figure}[t]
\includegraphics[width=8cm]{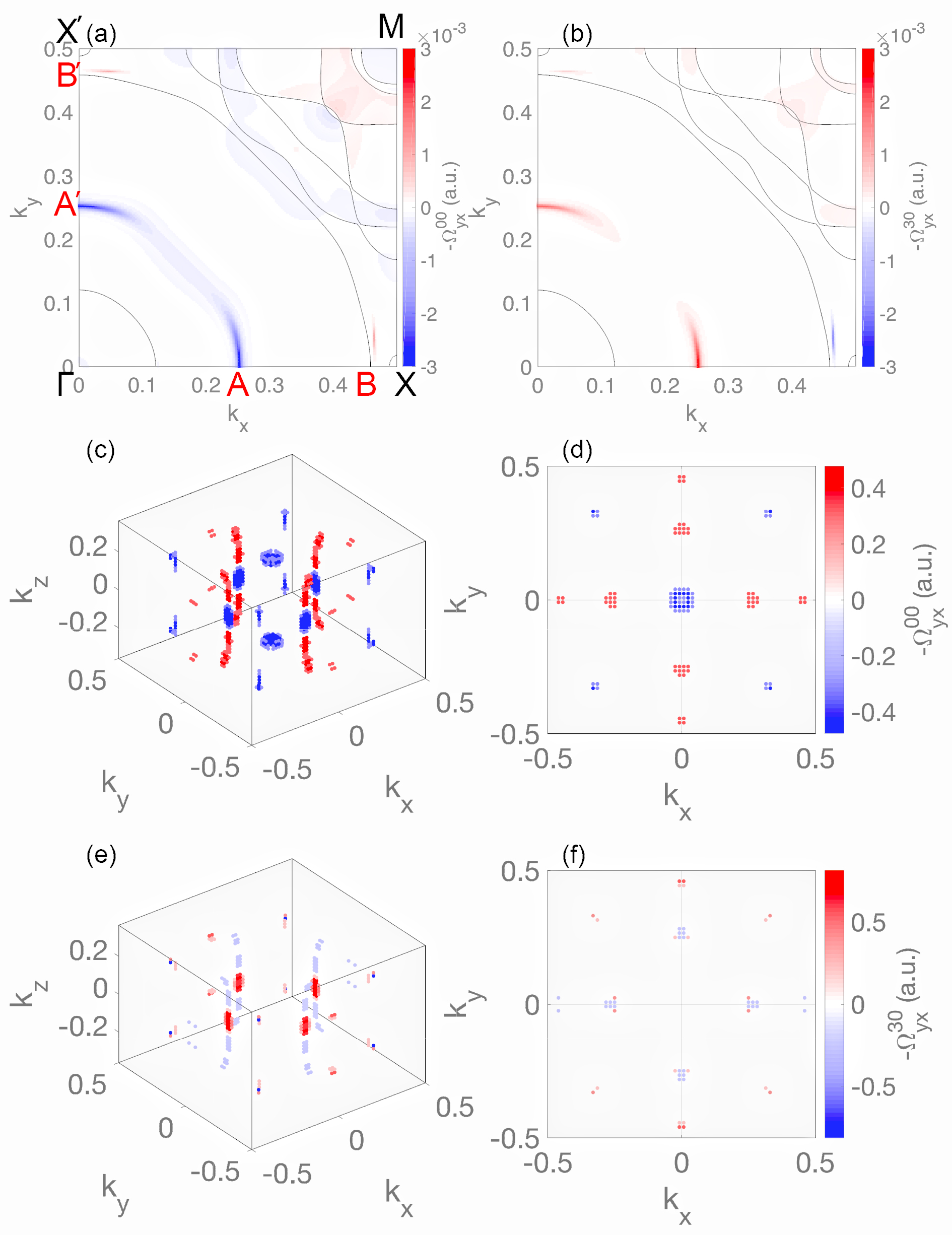}
\caption{\label{fig:2} (a,b) Berry curvature and spin Berry curvature mapping on (001) plane in a quarter of Brillouin zone. The black lines correspond Fermi contours on corresponding plane. Isometric view (c,e) and top view from (001) direction (d,f) of Berry and spin Berry curvature mapping inside the whole first Brillouin zone.  }
\end{figure}

Examples of class I appear at the A and A$^{\prime}$ points, which are located at the middle of the $\Gamma$-X and $\Gamma$-X$^{\prime}$ lines in Figs. \ref{fig:2}(a,b). At these points, a pair of bands with the same spin character (minority spin) form a gap at the Fermi level in the presence of SOC. $\Omega_{yx}^{00}$ and $\Omega_{yx}^{30}$ of the corresponding Fermi contours are highly correlated with each other, but with opposite signs, and are symmetric when applying the $C_{4v}$ rotation around the $z$-axis.
For these states, the electron spinor is oriented along the quantization axis (i.e. along the $z$-direction). Thus the off-diagonal components of the generalized velocity are negligible and $\Omega_{yx}^{00}$ and $\Omega_{yx}^{30}$ can be decomposed into those of majority spin part and minority spins part. The sign of $\Omega_{yx}^{00}$ follows the spinor state of the corresponding band whereas that of $\Omega_{yx}^{30}$ does not follow the polarization of the spinor state. 

A typical example of class II appears at the B and B$^{\prime}$ points where a pair of bands with opposite spin characters cross the Fermi level and a gap opens due to the SOC. The spin character of these states is therefore a mixture of majority and minority spins. 
In this case, a non-vanishing spin Berry curvature ($\Omega_{yx}^{30}$) appears with a positive sign at the B point (0.43,0,0) while it is negative at the crystallographically identical B$^{\prime}$ point (0,0.43,0). The Berry curvature ($\Omega_{yx}^{00}$), in contrast, is exactly the same for both B and B$^{\prime}$ points. To study the symmetry of the Berry and spin Berry curvatures with respect to the crystallographic symmetry, we present mapping of $\Omega_{yx}^{00}$ and $\Omega_{yx}^{30}$ inside the first BZ in Figs.~\ref{fig:2}(c, e): projections from the (001)-direction are shown in Figs.~\ref{fig:2}(d,f). As evident, we find $\Omega_{yx}^{00}$ and $\Omega_{yx}^{30}$ do not exhibit the same symmetry. Whereas $\Omega_{yx}^{00}$ possesses a $C_{4v}$ symmetry with respect to the magnetization direction ($z$-axis) following the symmetry of crystal structure, $\Omega_{yx}^{30}$ exhibits a reduced $C_{2v}$ symmetry, only reflecting the spatial inversion symmetry. 

\begin{figure}[t]
\includegraphics[width=8cm]{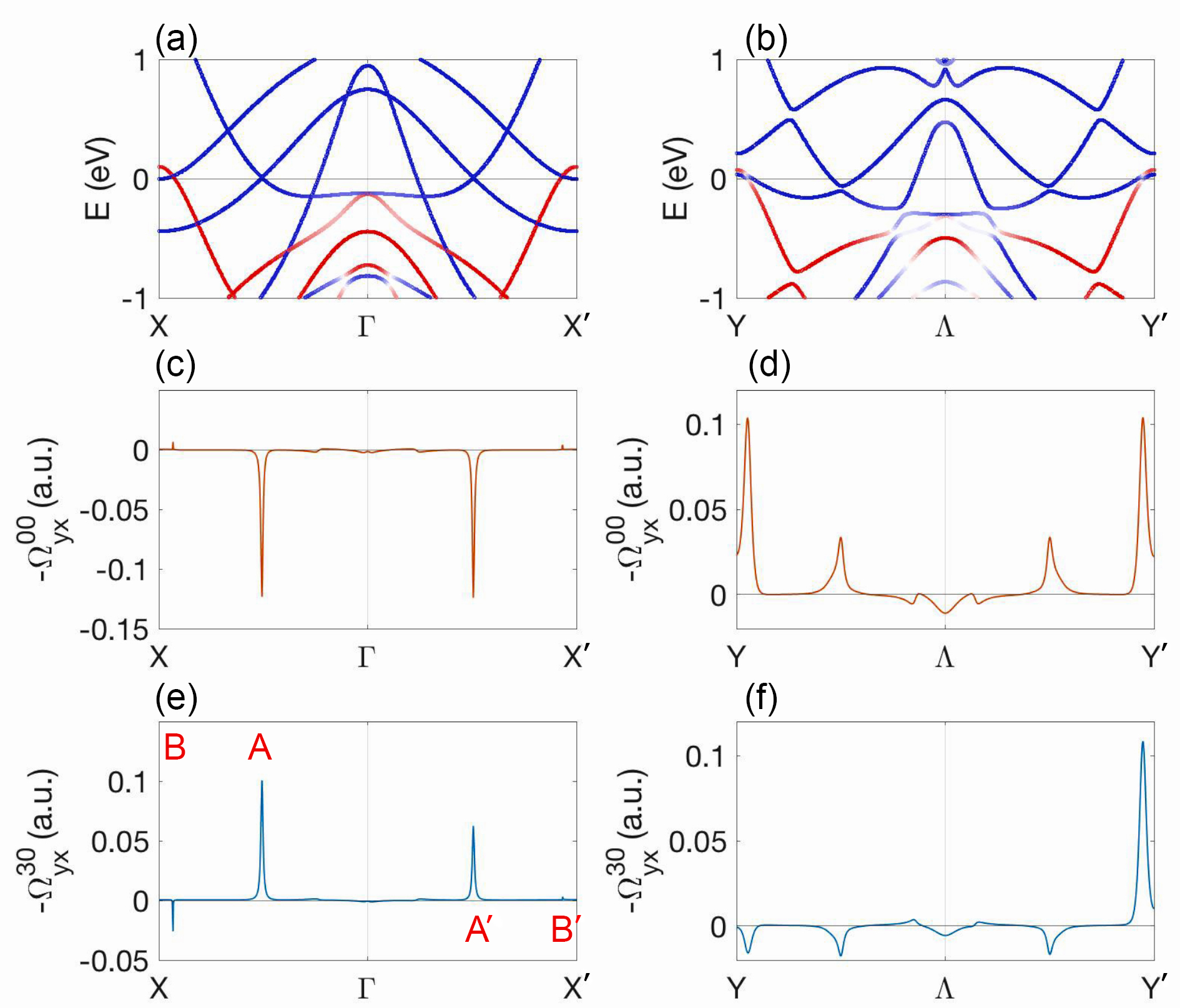}%
\caption{\label{fig:3} (a,b) The Band structure along selected symmetric k-paths X(0.5,0,0)-$\Gamma$(0,0,0)-X'(0,0.5,0) and Y(0,0.5,0.162)-$\Lambda$(0.5,0.5,0.162)-Y'(0.5,0,0.162). The bands are colored by its spin character which majority (red) and minority (blue). (c-f) Berry curvature and spin Berry curvature along selected k-paths.}
 \end{figure}

To quantitatively analyze the symmetry of the Berry and spin Berry curvatures, we show in Fig.~\ref{fig:3} $\Omega_{yx}^{00}$ and $\Omega_{yx}^{30}$ along selected symmetric path, X-$\Gamma$-X$^{\prime}$. (X$^{\prime}$ is the rotational symmetric point of X with respect to the $\Gamma$ point, see Fig.~\ref{fig:2}(e).) As expected, the band structure and spin character are exactly symmetric with respect to the $\Gamma$ point (Figs.~\ref{fig:3}(a,b)). For $\Omega_{yx}^{00}$ we confirm that all peaks and valleys hold identical values across the $\Gamma$ point(Figs.~\ref{fig:3}(c,d)). In contrary, the peaks in $\Omega_{yx}^{30}$ (Figs.~\ref{fig:3}(e,f)) across the $\Gamma$ point are different. 

For example, $\Omega_{yx}^{30}$ at the B$^{\prime}$ point is nearly zero whereas it is negative at the B point. The degree of asymmetry with respect to the absolute value of $\Omega_{yx}^{30}$ is larger at the B and B$^{\prime}$ points compared to that of the A and A$^{\prime}$ points. 
The small $\Omega_{yx}^{00}$ and $\Omega_{yx}^{30}$ at the B and B$^{\prime}$ points are due to the fact that the gap opening occurs at an energy level that is far from the Fermi level. To illustrate the asymmetry of $\Omega_{yx}^{30}$ more clearly, we plot another selected symmetric path, Y-$\Lambda$-Y$^{\prime}$, which is shifted by $k_z=0.162$. In this path, the Fermi level crosses exactly where the pair of bands with opposite spin character form the gap. The largest asymmetry is observed near the Y and Y$^{\prime}$ points, both in relative and absolute magnitude.

From these results, we find that the largest asymmetry of $\Omega_{yx}^{30}$ within the crystallographically equivalent points appears at locations where a pair of band with opposite spin characters forms a gapped state via the SOC (i.e. class II). Here the spin of the Bloch states are nearly degenerate. For pair of bands with the same spin character, the asymmetry of $\Omega_{yx}^{30}$ at crystallographically equivalent points is almost negligible.

 \begin{figure}[t]
 \includegraphics[width=8cm]{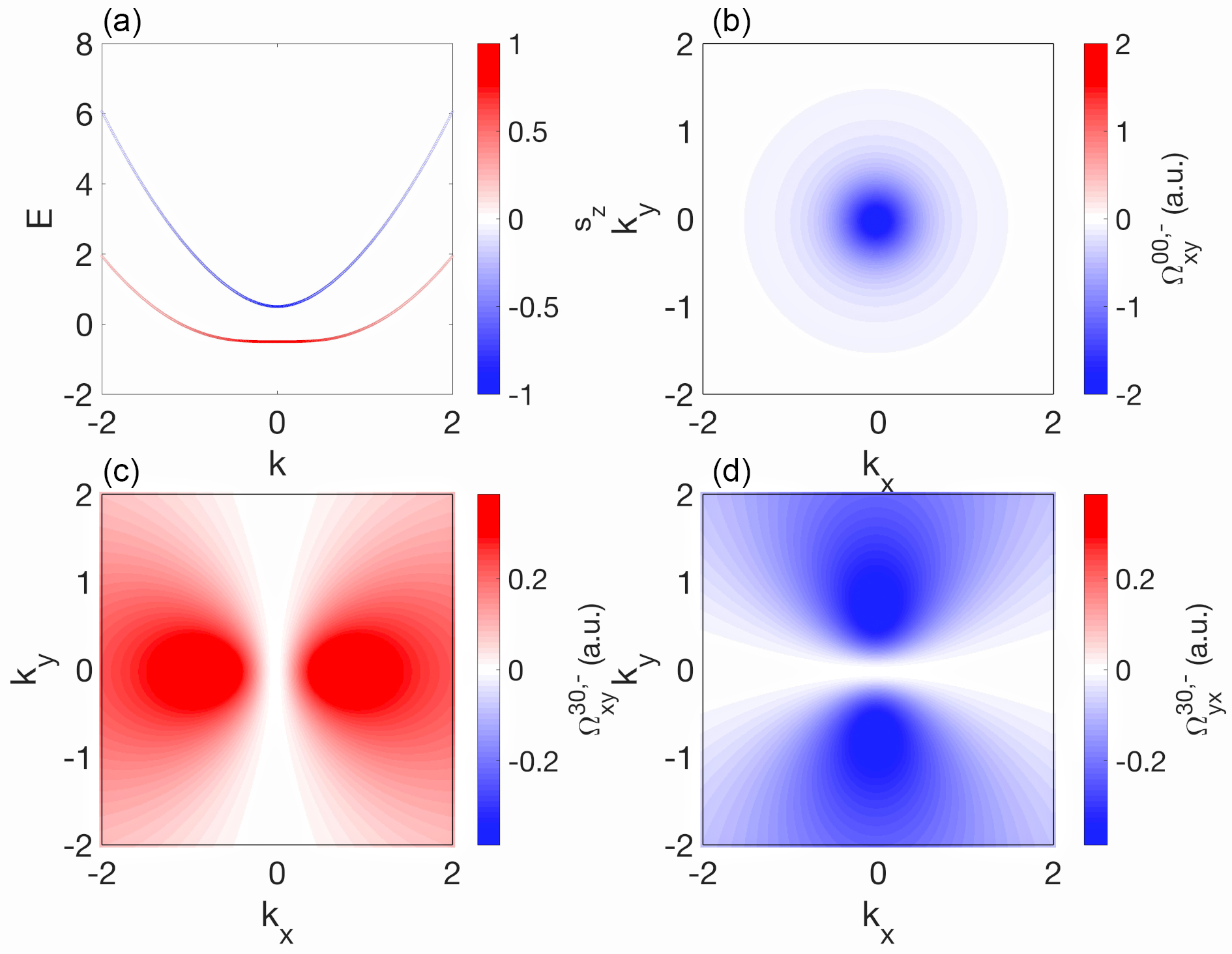}%
 \caption{\label{fig:4} (a) Band dispersion of model Hamiltonian whose color is coded with spin character. (b) Berry curvature $\Omega_{xy}^{0,-}$ of lower band.  (c,d) spin Berry curvature $\Omega^{30,-}_{xy}$/$\Omega^{30,-}_{yx}$ of lower band. The parameters are chosen as, $\alpha=1$ $\beta=0.5$.} 
 \end{figure}
 
We note that the reduction of the spacial symmetry of the spin Berry curvature with respect to the symmetry of the crystal (and the Berry curvature) is a general feature. To illustrate this, we use the following model Hamiltonian (i.e. the so-called Rashba Hamiltonian) widely used to characterize electron transport in ferromagnets with spin orbit interaction\cite{nagaosa2010anomalous}:
\begin{equation}
\label{eq:6}
\mathcal{H} = k_x^2+k_y^2 + \alpha \Big( \sigma_1 k_y - \sigma_2 k_x \Big) + \beta \sigma_3
\end{equation}
$\alpha,\beta$ are parameters controlling the SOC and exchange splitting, respectively. We employ the simplest model Hamilton representing a ferromagnet, which contains spin orbit coupling (Rashba-like) and exchange splitting. This system can simulate the situation of Type II in our first principle calculations. The energy dispersion relation of the model Hamiltonian is shown in Fig. \ref{fig:4}(a). Due to the exchange splitting, there are two bands which we refer to as the lower $(-)$ (red line) and upper $(+)$ (blue line) bands. This model thus can be considered as a representative case of class II; the gap is formed with a pair of bands with opposite spins. The Berry and spin Berry curvatures 
are calculated using the Kubo formula for the $(-)$ and $(+)$ bands as,
\begin{equation}
 \Omega_{xy}^{00,\pm}=\pm \frac{\alpha^2 \beta}{2 \lambda^3 (\bm{k})}, \ \Omega_{xy}^{30,\pm}=\mp \frac{\alpha^2 k^2_x}{2 \lambda^3 (\bm{k})}, \ \Omega_{yx}^{30,\pm}=\pm \frac{\alpha^2 k^2_y}{2 \lambda^3 (\bm{k})}
 \label{eq:7}
 \end{equation}
 where $\lambda (\bm{k}) =\sqrt{\alpha^2 \bm{k}^2+\beta^2}$.
 
The calculated Berry and spin Berry curvatures of the $(-)$ band are presented in Fig.~\ref{fig:4}(b-d). Note that the $C_{4v}$ rotation symmetry is a subgroup of the spherical symmetry group. Since the spin character of the two bands is opposite, the Berry curvature is $C_{4v}$-symmetric whereas the spin Berry curvature reduces to $C_{2v}$-symmetry. 

To explicitly show the symmetry of the Berry and spin Berry curvatures, we rewrite the Kubo formula as the following,
 \begin{widetext}
\begin{equation}
\Omega_{ij}^{\alpha 0}(\bm{k}) = -\frac{1}{2} \sum_{n' \neq n} \Big[ f(\epsilon_n( \bm{k} )) -f(\epsilon_n'( \bm{k} )) \Big] \times \frac{ \text{Im} \Big[ [ v_i^{\alpha} ]_{n,n'}( \bm{k} )  [  v_j^{0} ]_{n',n}( \bm{k} ) -[ v_j^{0} ]_{n,n'}( \bm{k} ) [ v_i^{\alpha} ]_{n',n}( \bm{k} )  \Big] }{\Big(\epsilon_{n}( \bm{k} ) - \epsilon_{n'}( \bm{k} ) \Big)^2}
 \label{eq:3}
\end{equation}
\end{widetext}
where $ [ v_i^{\alpha} ]_{n',n}( \bm{k} )= \bra{ \bm{k} , n'} \hat{v}_i^{\alpha} \ket{ \bm{k} , n}$ is the matrix element of the generalized velocity operator $ \hat{v}_i^{\alpha}$ evaluated with a Bloch state with a wave vector $\bm{k}$. For a system having a $C_{4v}$ symmetry around the $z$-axis, the global rotation operator on a spinor reads,
\begin{equation}
\ket{ \mathbf{\Lambda} \bm{k},n} = D(\mathbf{\Lambda}) \ket{\bm{k},n}
\end{equation}
where $D( \mathbf{\Lambda} )=\exp(i \vec{\theta} \cdot \vec{\sigma}/2)$. The Pauli matrix $\vec{\sigma}$ is the generator of $SU(2)$ group. We define $\mathbf{\Lambda}=\exp(i \vec{\theta} \cdot \vec{J})$, where $\vec{J}$ is the generator of $SO(3)$ group.

Since the crystal field of CoPt preserves a $C_{4v}$ rotation symmetry around the $z$-axis, the band dispersion relation and the Bloch states are both invariant under $C_{4v}$ rotation. However, the generalized velocity matrix element is not invariant. 
Using the identities $D^{\dagger}(\mathbf{\Lambda}) \sigma_{i} D(\mathbf{\Lambda})=\Lambda_{ij} \sigma_{j}$  and $\partial_{\mathbf{\Lambda} \bm{k}} =\dfrac{\partial_{\mathbf{\Lambda} \bm{k}}}{\partial_{\bm{k}}} \partial_{\bm{k}}=\mathbf{\Lambda}  \partial_{\bm{k}}$, the generalized velocity operators can be written as:
\begin{eqnarray}
[ v_i^{0} ]_{n',n}( \mathbf{\Lambda} \bm{k}) &=& \Lambda_{ij} [  v_j^{0} ]_{n',n}(\bm{k}) \notag \\
{}[ v_i^{l} ]_{n',n} ( \mathbf{\Lambda} \bm{k}) &=& \Lambda_{ij} \Lambda_{lp} [  v_j^{p} ]_{n',n}(\bm{k})
\label{eqn:4}
\end{eqnarray}
Under $C_{4v}$ rotation, the Berry and spin Berry curvatures (Eq.~\ref{eq:3}) have the following relations:
\begin{eqnarray}
\Omega_{ij}^{00}(\mathbf{\Lambda} \bm{k} ) &=& \Lambda_{ir} \Lambda_{js} \Omega_{rs}^{00} (\bm{k})  \notag \\
\Omega_{ij}^{l0}(\mathbf{\Lambda} \bm{k} ) &=& \Lambda_{ir} \Lambda_{js} \Lambda_{lp} \Omega_{rs}^{p 0} (\bm{k})
\label{eqn:5}
\end{eqnarray}

The Berry curvature preserves the $C_{4v}$ symmetry, i.e. $\Omega_{yx}^{00}(\mathbf{\Lambda} \bm{k}) =-\Omega_{xy}^{00}( \bm{k})=\Omega_{yx}^{00}( \bm{k})$ when $\displaystyle{\theta=(0,0,\frac{\pi}{2})}$ is substituted in Eq. (\ref{eqn:5}). The last equality can be directly confirmed from Eq.~(\ref{eq:3}) where the spatial indices of the Berry curvature can be exchanged anti-symmetrically. Similarly, the spin Berry curvature satisfies $\Omega_{yx}^{30}(\mathbf{\Lambda} \bm{k}) =-\Omega_{xy}^{30}( \bm{k}) = \Omega_{yx}^{03}( \bm{k}).$ In the last equality, note the exchange of the spatial indices in the spin Berry curvature defined in Eq.~(\ref{eq:3}) leads to exchanging the spin indices as well. 
The spin Berry curvature, however, breaks the $C_{4v}$ symmetry since $\Omega_{yx}^{03}$ is not necessarily equal to $\Omega_{yx}^{30}$ at all $k$ points, i.e. $\Omega_{yx}^{30}(\mathbf{\Lambda} \bm{k}) \neq \Omega_{yx}^{30} ( \bm{k})$. 

For the case of class I, the Bloch states at a certain $k$ point have their spin directions aligned along the quantization axis (i.e. along the $z$ axis), and therefore the corresponding matrix element of the generalized velocity operator $[  v_i^{0} ]_{n',n}( \bm{k} )$ and $[  v_i^{3} ]_{n',n}( \bm{k} )$ are identical up to a factor of $\pm 1$. Thus in such case the spin Berry curvature follows the symmetry of the Berry curvature and the band structure. In contrast, for class II, the Bloch states are composed of both the majority and minority spin characters, and thus the off-diagonal elements of the generalized velocity operator become non-negligible, contributing to the difference of the velocity operator and the spin current operator. Such difference in the two operators give rise to the distinct symmetry difference of the Berry and spin Berry curvatures.

As ferromagnetic systems typically possess a large exchange splitting, the Hall current is often modeled using the two current model, i.e. the Hall current consists of two independent channels formed by the majority and minority spins. Using a simplified two current model, the anomalous Hall effect and the spin Hall effect are related by the spin polarization at the Fermi level $P$, that is, $\sigma_{AH} = P \sigma_{SH}$. This simple scaling does not hold for intrinsic contribution of SHC and AHC in CoPt, as we find the spin polarization $P=\sigma^{00}_{yx}/ \sigma^{30}_{yx}$ is nearly zero from the calculated Hall conductivities ($\sigma_{AH}=-3$ S/cm and $\sigma_{SH}=787$ S/cm). This is because pairs of bands with opposite spin characters (class II) are ubiquitous inside the Brillouin zone near the Fermi level in ferromagnets. Under such circumstance, the Berry curvature and the spin Berry curvature are not simply related by the spin polarization of bands due to the non-zero off diagonal elements of the generalized velocity operator. With such states contributing to total conductivity, the AHC does not scale with the SHC where the scaling factor is the spin polarization at the Fermi level. Interestingly, recent experimental reports show that indeed such simple scaling does not hold for the intrinsic and extrinsic contributions to the AHC and SHC in 3d ferromagnets\cite{omori2018relation}.

In conclusion, we have studied the L1$_0$-CoPt to investigate the Berry curvature and the spin Berry curvature in a ferromagnetic system. We find that the symmetry of the spin Berry curvature is reduced from that of the Berry curvature and the band structure. Based on the Kubo formula and model Hamiltonian analyses, the reduced symmetry of the spin Berry curvature originates from band crossing points in the reciprocal space where the Bloch states have opposite spin characters. The presence of such state not only influences the symmetry of the spin Berry curvature but also alters the scaling relation between the anomalous Hall conductivity and the spin Hall conductivity. These results suggest that the two current model is too simplified to derive the relation between the  anomalous Hall and spin Hall conductivities in ferromagnets.

\section*{Appendix}
\subsection{DFT calculations}
DFT calculations are performed using the full-potential linearized augmented-plane-wave method (FLAPW) with generalized gradient approximation (GGA) for exchange correlation. LAPW functions have a cutoff, $|k+G| \leqslant 3.9 \text{ a.u.}^{-1} $. Muffin-tin (MT) radius are taken to be 2.2 and 2.4 bohrs for Co and Pt, respectively. The angular momentum expansion inside the MT spheres is truncated at $l=8$ for the wave functions, charge and spin densities, and potential. The size of $k$-point mesh is selected as $16 \times 16 \times 16$ for obtaining self-consistent charge and spin densities. 
 
 To check the accuracy of the calculations of the AHC and SHC, we extend the size of $k$-point mesh up to $70 \times 70 \times 70$ with a total 343,000 special $k$ points inside the first BZ. As shown in Fig. \ref{fig:1}(c), changes in the integrated SHC is less than $5\%$. The calculated spin magnetic moment of Co and Pt are 1.76 and 0.40 $\mu_{B}$ respectively, which show good agreement with previous calculations, confirming the reliability of the calculations. 
With zero-temperature assumption, the Fermi distribution reduces to a step function. 
 


\begin{acknowledgments}
\section*{Acknowledgements}
This work was partly supported by JSPS Grant-in-Aid for Specially Promoted Research (15H05702) and the Center of Spintronics Research Network of Japan. Work at Mie University was supported by JSPS KAKENHI Grant Number 16K05415.
\end{acknowledgments}

\bibliography{CoPt_011019}

\end{document}